\documentclass[reprint,superscriptaddress,amsmath,amssymb,floatfix,pra]{revtex4-2}
\usepackage{amsmath,amssymb,mathtools}
\usepackage{graphicx}
\usepackage{bm}
\usepackage[colorlinks, citecolor=blue, linkcolor=red]{hyperref}
\usepackage{braket}
\newcommand{\diag}{\operatorname{diag}}
\begin{document}
	\title{Open quantum dynamics with singularities: Master equations and degree of non-Markovianity}
	\author{Abhaya S. Hegde}
	\email{abhayhegde16@iisertvm.ac.in}
	\author{K. P. Athulya}
	\author{Vijay Pathak}
	\affiliation{%
		School of Physics, Indian Institute of Science Education and Research, Thiruvananthapuram, Vithura, Kerala 695551, India
	}%
	\author{Jyrki Piilo}
	\affiliation{%
	    Turku Centre for Quantum Physics, Department of Physics and Astronomy, University of Turku, FI-20014, Turun yliopisto, Finland
	}%
	\affiliation{%
		Laboratory of Quantum Optics, Department of Physics and Astronomy, University of Turku, FI-20014, Turun yliopisto, Finland
	}%
	\author{Anil Shaji}
	\email{shaji@iisertvm.ac.in}
	\affiliation{%
		School of Physics, Indian Institute of Science Education and Research, Thiruvananthapuram, Vithura, Kerala 695551, India
	}%
	\date{\today}
	
	\begin{abstract}
		Master equations describing open quantum dynamics are typically first-order differential equations. When such dynamics brings the trajectories in state space of more than one initial state to the same point at finite instants in time, the generator of the corresponding master equation becomes singular while the dynamical map becomes non-invertible. The first-order, time-local, homogeneous master equations then fail to describe the dynamics beyond the singular point. Retaining time-locality in the master equation necessitates a reformulation in terms of higher-order differential equations. We formulate a method to eliminate the divergent behavior of the generator by using a combination of higher-order derivatives of the generator with suitable weights and illustrate it with several examples. We also present a detailed study of the central spin model and we propose the average rate of information inflow in non-Markovian processes as a quantity that captures a different aspect of non-Markovian dynamics.
	\end{abstract}
	
	\maketitle
	
	\section{\label{sec:intro}Introduction}
	Almost all realistic quantum systems are open systems with their dynamics determined by interactions with the environment also. Although the evolution of the system in the presence of its environment does not follow unitary dynamics, the combined evolution of the system and environment is unitary in nature. The reduced dynamics of the system of interest is then obtained by tracing over the environmental degrees of freedom from the time-evolved combined density matrix as $\rho_{S}(t)=\text{Tr}_{E}[\rho_{SE}(t)]$. The reduced system dynamics induced by the joint evolution of the system and its environment can be modeled by a dynamical map given by $\rho_{S}(t)=\mathcal{E}_{t} \rho_{S}(0)$~\cite{Sudarshan1961, Choi1972, Choi1975, Erling1963}. While the dynamical maps describe changes in the state of the open system across finite-time intervals akin to the unitary time-evolution operator for closed systems, continuous-time description of open quantum evolution is typically formulated in terms of quantum master equations~\cite{BreuerBook, Breuer2009}. Open quantum systems endowed with a large separation in timescales of the system and environment are modeled using the Markov approximation and their dynamics is described by a Markovian master equation. The quantum master equation under the Markov approximation can be written in the Gorini-Kossakowski-Sudarshan-Lindblad (GKSL) form that corresponds to completely positive and trace preserving open quantum system dynamics~\cite{GKS1976, Lindblad1976}.
	
	There are processes for which the Markovian approximation is not valid and we have to turn to non-Markovian dynamics. Time-dependent, local-in-time, master equations of GKSL form can be formulated for the non-Markovian case as well~\cite{Breuer1999, Silvan2016, Sabrina2014, Rivas_2012}. In this paper, we present several physically realizable non-Markovian cases for which forcing the description of the system dynamics into a time-local master equation leads to a singular generator. The propagation of states after the singularity cannot be done formally using the time-local master equation. Motivated by the rapid developments in the ability to study open quantum dynamics experimentally, we address this gap in the formalism and in the process also propose a minor but useful modification to one of the standard ways of quantifying information back flow and non-Markovianity.
	
	We investigate how processes in which the trajectories of distinct states diverge after a singularity can be mathematically described. Note that the trajectories we consider in the following are in the space of all possible quantum states of the system of interest. A suitable parametrization of the state space, for instance, with the Bloch ball of states of a single qubit, will allow us to visualize these trajectories as well. We see how a general master equation for such dynamics that holds true for all time can be constructed in certain cases. Specifically, we propose higher-order master equations to weed out the singularities in a manner that their solutions reduce to that of the traditional first-order equation at all other points. The proposed higher-order equations naturally take care of propagating the state through the singularities. Dynamics with singular points are typically non-Markovian. Different approaches to characterize the non-Markovianity resulting from the divergent behavior of generators were studied in~\cite{Hall2007, Hall2014, DivisibilityPRA2011, DivisibilityPRA2019, DivisibilityPRL2018} and a measure to characterize the nature and degree of the singularity was proposed in~\cite{Hou2012}. Depending on the nature of the singularity in the generator of the first-order master equation, we arrive at different forms of equivalent higher-order master equations that avoid the singular behavior. We are interested in exploring the connection, if any, between the nature of the singularity and the nature of the non-Markovianity in the system. This however requires a comparison of the degree of non-Markovianity in different processes. There are several proposed measures of non-Markovianity available in the literature~\cite{Li2018,Li2019,Li2020,Wolf2008,
	RHP2010, Vasile2011,
	Lu2010, Lorenzo2013, Alipour2012, breuer_measure_2009, Laine2010, Liu2013,ColloquiumNM, Daniel2017, rivas_huelga_plenio_2014}, but they do not typically allow for a direct comparison between processes as explained later on. The characterization of the singularity in~\cite{Hou2012} also is not suitable for comparison of different processes. To circumvent these difficulties, we introduce a quantity to capture the persistence of information inflow which, in turn can lead to meaningful comparison of different non-Markovian processes. In addition to using this quantity to compare the singular processes, we extend its applicability and demonstrate its utility in comparing generic non-Markovian processes as well.
	
	This paper is structured as follows. In Sec.~\ref{sec:def}, we introduce the relevant definitions and the problem. We reinforce the issues of singular dynamics with an illustrative example in Sec.~\ref{sec:spin_model}. A discussion on possible avenues to resolve the issue is presented in Sec.~\ref{sec:met}. In Sec.~\ref{sec:Solution}, we apply our results to the example presented in Sec.~\ref{sec:spin_model}. We comment on different classes of examples in Sec.~\ref{sec:more_ex} using our methods. A new quantity that enables comparison of the observed non-Markovianity in different processes is proposed in Sec.~\ref{sec:measures}.  Section~\ref{sec:disc} contains a brief discussion and our conclusion.
	
	\section{\label{sec:def}Dynamical Maps and Master Equations}
	The dynamics of an open quantum system that is initially in a product state with its environment can be expressed in terms of the completely positive and trace preserving (CPTP) dynamical maps $\mathcal{E}_t$. The open system we will be considering is a single qubit. Using the left-right vectorization formalism~\cite{Milz_2017} to write the equations of motion for the open dynamics of the qubit, we represent its quantum states, $\rho_t$, by real vectors and the quantum dynamical maps $\mathcal{E}_t$ as real four-dimensional matrices. Since the Hilbert space associated with a qubit is a subset of the four-dimensional linear space of Hermitian qubit operators, it follows that any quantum state can be written as $\rho = (\mathbb{I}+\vec{r} \cdot \vec{\sigma})/2$, where $|\vec{r}| \leq 1$ and $\vec{\sigma} = (\sigma_x, \sigma_y, \sigma_z)$ is a vector of Pauli operators. The condition $|\vec{r}| \leq 1$ enforces positivity of $\rho$ and the states of the qubit can be represented as points in the Bloch sphere. The vector $(1, \vec{r})$ furnishes the real, four-dimensional representation of the quantum state. The affine form of $\mathcal{E}_t$ acting on the state $(1, \vec{r})$ is
	\begin{equation}
	\label{map}
	\mathcal{E}_t = \begin{pmatrix} 1 & \vec{0} \\ \vec{s} & T \end{pmatrix},
	\end{equation} 
	with $\vec{s}$ a translation vector and $T$ a real three-dimensional matrix. The Bloch sphere vectors transform as $\vec{r}' \equiv \mathcal{E}_t(\vec{r}) = T\vec{r}+\vec{s}$. 
	
	The state of the system at time $t$ is given by the dynamical map as
	\begin{equation}
	\label{dyn_map}
	\rho_t = \mathcal{E}_t [\rho_0],
	\end{equation}
	with $\rho_0 \equiv \rho_{t=0}$. When $\mathcal{E}_t$ is an invertible map, one finds its time-local generator as 
	\begin{equation}
	\label{generator_map}
	\mathcal{L}_t = \dot{\mathcal{E}}_{t}^{\mathstrut}\mathcal{E}_{t}^{-1}.
	\end{equation}
	Assuming the semigroup property $\mathcal{E}_{t+s} = \mathcal{E}_t\mathcal{E}_s$, we can write a time-local master equation $\dot{\rho}_t = \mathcal{L}_t [\rho_t]$ in the well-known GKSL form (choosing $\hbar = 1$)
	\begin{equation}
	\label{GKSLeq}
	\dot{\rho}_{t} = -i[H, \rho_t] + \sum_{i=1}^{3}\gamma_{i}\bigg(L_{i}^{\mathstrut} \rho_{t}^{\mathstrut} L_{i}^\dagger - \frac{1}{2}\Big\{L_{i}^\dagger L_{i}^{\mathstrut}, \rho_{t}^{\mathstrut}\Big\}\bigg)
	\end{equation}
	where $\text{tr}(L_i)=0$ and $\text{tr}(L_iL_j)=\text{tr}(L_jL_i)=\delta_{ij}$. In other words, the Lindblad operators $L_i$ are traceless and orthonormal. The dynamics described by the semigroup master equation is Markovian. The Markovian master equation may be generalized by introducing time-dependent Lindblad-like operators and time-dependent decay rates $\gamma_i(t)$ in Eq.~\eqref{GKSLeq}. This results in a GKSL form for generators $\mathcal{L}_t$ of open dynamics that are not Markovian in general,
	\begin{align}
	\label{quasi_GKSL}
	\dot{\rho}_t = & -i[H(t), \rho_t] \nonumber \\ 
	& + \sum_{i=1}^{3} \gamma_i(t) \left[L_{i}^{\mathstrut}(t) \rho_{t}^{\mathstrut} L_{i}^{\dagger}(t) - \frac{1}{2}\left\{L_{i}^{\dagger}(t) L_{i}^{\mathstrut}(t), \rho_{t}^{\mathstrut}\right\}\right].
	\end{align}
	The presence of negative rates $\gamma_i(t) < 0$ for some $i$ and $t$ can be regarded as non-Markovian behavior~\cite{Hall2014, ColloquiumNM, Daniel2017, rivas_huelga_plenio_2014}.
	
	Since the time-local master equation is first-order in time, knowing the state at time $t$ allows one to uniquely determine the state at all later times $t' > t$. In particular, it follows that if the trajectories of two states $\rho_1(t)$ and $\rho_2(t)$ intersect at some time $t = t_{c}$, i.e., $\rho_1(t_c) = \rho_2(t_c)$, the trajectories will move together for all subsequent times, i.e., $\rho_{1}(t') = \rho_{2}(t')$ for $t' > t_{c}$. Since any such merging of trajectories is irreversible, the dynamical map in Eq.~\eqref{dyn_map} becomes noninvertible in all such cases and thus the generator as defined in Eq.~\eqref{generator_map} ceases to exist. However, there are several examples of physically valid processes in which the trajectories of multiple states converge at distinct points in time and then are again separate for $t > t_{c}$. Moreover, the trajectories of qubit dynamics visualized on the Bloch sphere for all such processes are analytic, even at those instants of time when the inverse dynamical map does not exist. Clearly, the first-order equation fails to describe the dynamics of these processes. We illustrate such a process using the central spin model in the next section and then propose a way of describing such dynamics using higher-order differential equations. 

	\section{\label{sec:spin_model}Example: Central Spin Model}
	
	To illustrate the problem at hand, we examine here a central spin model used to simulate the interaction of a single electron spin confined to a quantum dot with a bath of nuclear spins~\cite{Breuer2004}. Consider a bath consisting of $N$ spin-$\frac{1}{2}$ particles coupled to a central spin-$\frac{1}{2}$ particle. The interaction Hamiltonian is 
	\begin{equation}
	\label{spin_ham}
	H = \sum_{k=1}^{N}A_k \sigma_z \otimes \sigma_z^{(k)}, \quad A_k = \frac{A}{\sqrt{N}}
	\end{equation}
	such that each spin in the bath is interacting with the central spin via the Pauli $\sigma_z$ operator. Note that we have scaled the coupling constant appearing in the Hamiltonian by a factor of $1/\sqrt{N}$ that will keep the total interaction energy between the central spin and the ones around constant irrespective of $N$. We will see later on that this choice is required if we are to compare different non-Markovian processes. We begin with an initial product state for the total system of $N+1$ particles such as, $\eta_{0} = \rho_{0} \otimes \mathbb{I}/2^N$. The final state of the central spin after tracing out the bath of $N$ surrounding spins at time $t$ is
	\begin{align}
	\rho_{t} &= \text{Tr}_{\text{E}}\left(e^{-iHt}\eta_{0}e^{iHt}\right) \nonumber \\
	& =
	\begin{pmatrix}
	\rho_{11} & \cos^N\left(\frac{2At}{\sqrt{N}}\right)\rho_{12} \\
	\cos^N\left(\frac{2At}{\sqrt{N}}\right)\rho_{21} &\rho_{22}
	\end{pmatrix}
	\label{sys_state}
	\end{align}
	with $\rho_{ij}$ for $i,j = \{1, 2\}$ as the elements of $\rho_{0}$. The master equation typically used to describe this process is~\cite{Hou2012}
	\begin{equation}
	\label{spin_mod}
	\dot{\rho}_{t} = A\sqrt{N}\tan\left(\frac{2At}{\sqrt{N}}\right)\left(\sigma_z\rho_{t}\sigma_z-\rho_{t}\right).
	\end{equation}
	
\begin{figure}[!htb]
	    	\includegraphics[width=0.48\columnwidth]{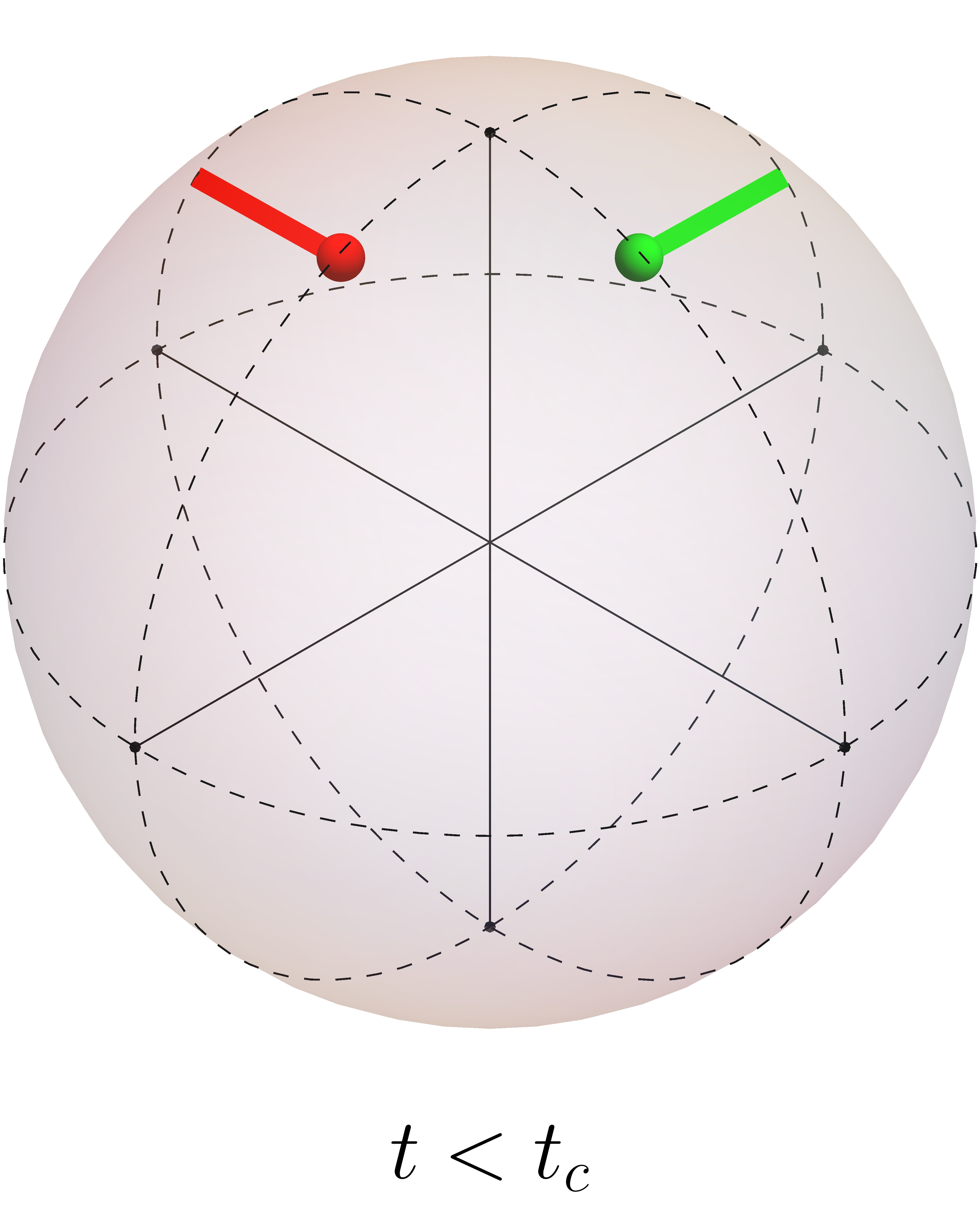}
	    	\includegraphics[width=0.48\columnwidth]{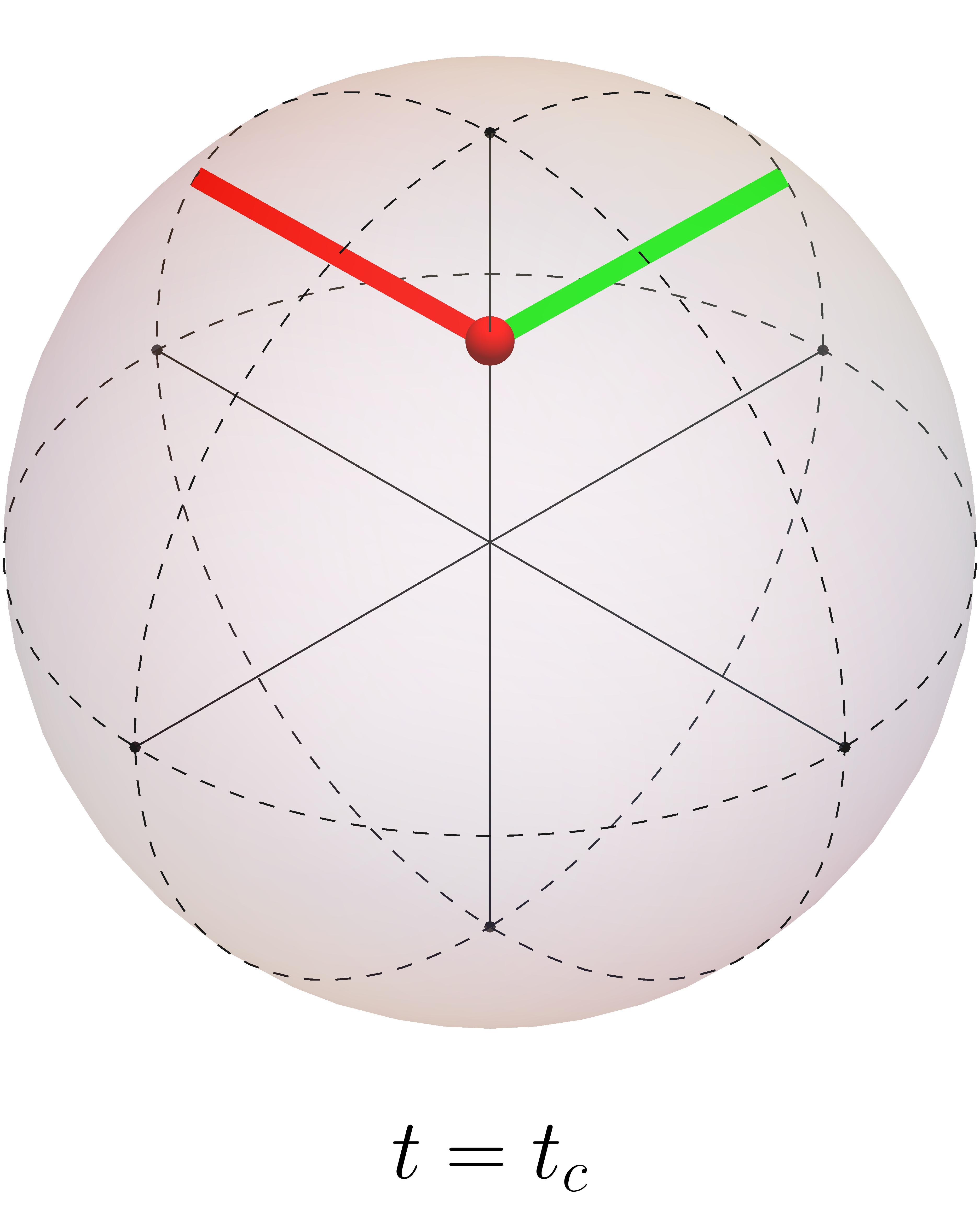}
	    	\includegraphics[width=0.48\columnwidth]{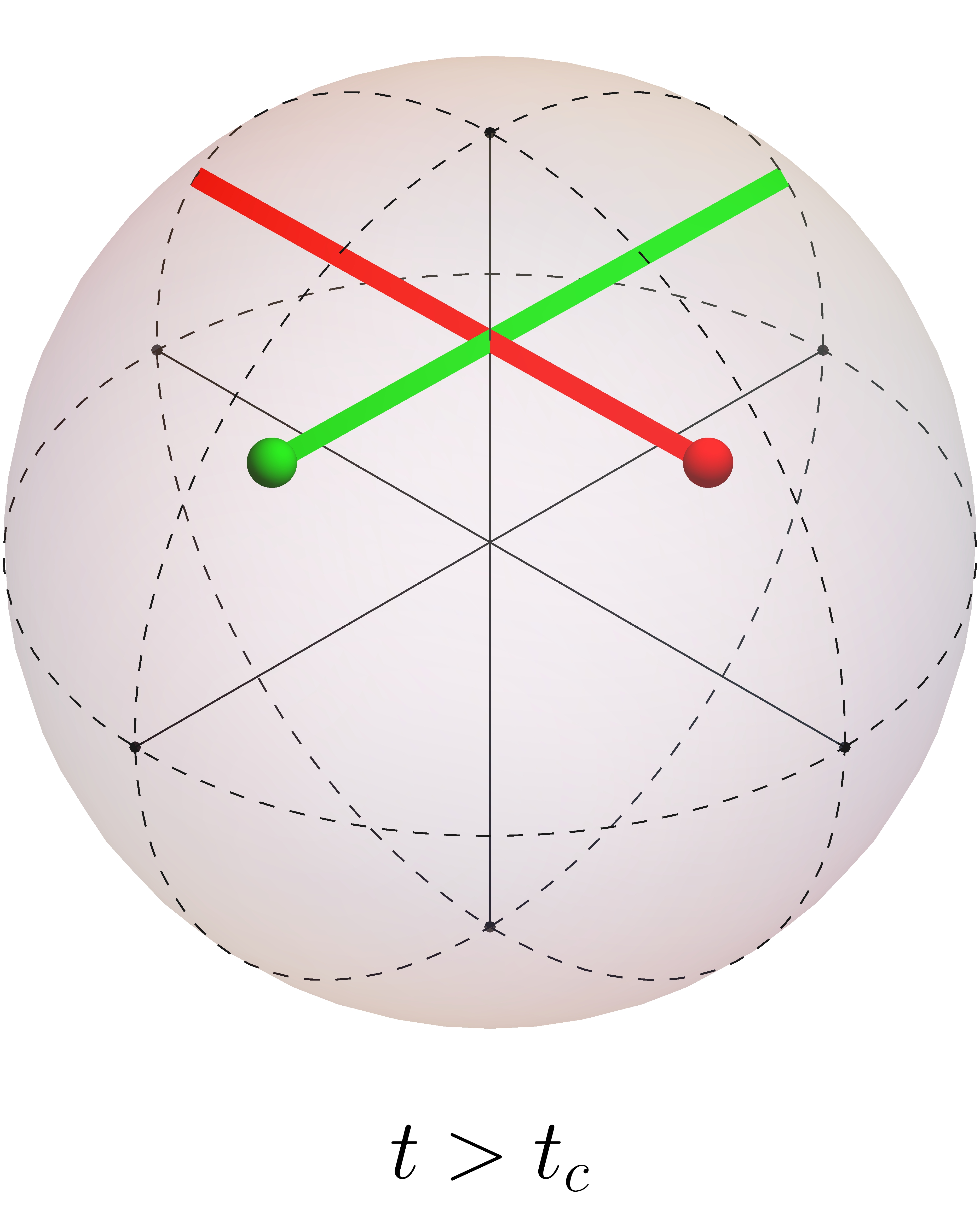}
	    	\label{fig:lab3}
	    	\caption{\label{fig:sing_behav} Trajectories of two initially distinct states are shown at different times on the Bloch sphere. The figure corresponds to the evolution given by the Hamiltonian in Eq.~\eqref{spin_ham} with $N=1$.}
	    \end{figure}

	The rate appearing in the equation above is proportional to $\tan(t)$ and the equation is singular for all $t = \sqrt{N}(2k+1)\pi/4A$ for $k = 0,\, 1,\, 2,\, \cdots$. However, this model is known to be exactly solvable for all $N$~\cite{Breuer2004}. Moreover, it is easy to see that the dynamical map corresponding to this process,
	\begin{equation}
	\label{spin_map}
	\mathcal{E}_{t}^{\text{spin}} = \diag\left(1,\, \cos^{N}\bigg(\frac{2At}{\sqrt{N}}\bigg), \, \cos^{N}\bigg(\frac{2At}{\sqrt{N}}\bigg), \, 1\right)
	\end{equation}
	is a well-defined diagonal matrix for all $t$. The trajectories of a pair of initial states of the central spin are plotted on the Bloch sphere in Fig.~\ref{fig:sing_behav}. We see that the two trajectories intersect at $t=t_c$ and the inverse map, ${\mathcal E}_t^{-1}$ becomes one-to-many and singular at this point. The master equation~\eqref{spin_mod} fails to describe the observed trajectory beyond this (first) singular point since beyond $t_c$ the first-order differential equation yields identical evolution for both intersecting trajectories. The dynamical map outputs the correct final state for all times nevertheless and yields diverging trajectories after $t=t_c$ as shown in the figure. The failure of the master equation to predict the evolution beyond $t_c$ prompts us to explore the existence of an alternate differential equation that is consistent with the dynamics given by the map while at the same time, does not exhibit singularities.
	
	\section{\label{sec:met}Higher Order Master Equations}
	
	Since any nondiagonal dynamical map can be made diagonal by a suitable choice of operator basis~\cite{nielsen2002quantum}, we explore the case of a general diagonal map. For the sake of simplicity we will stick to unital maps for which $\vec{s} = 0$ in Eq.~\eqref{map}. We point out that our arguments can also be extended to non-unital maps in a straightforward manner, as shown in examples later (see Sec. \ref{subsec:non_uni} below). Choosing the affine map in Eq.~\eqref{map} as a diagonal matrix that describes the transformations of a state in each subspace, we define
	\begin{equation}
	\label{map_form}
	T = \diag{(f_{x}(t),\, f_{y}(t),\, f_{z}(t))}.
	\end{equation}
	Writing the initial state in the vectorized form $\rho_0 = (1, x, y, z)^T$, the action of this unital map corresponds to the master equation:
	\begin{align*}
	\frac{d \rho}{dt} &= \dot{\mathcal{E}}_t \rho_0 \\
	& = {\left(
		\begin{array}{cccc}
		1 & 0 & 0 & 0 \\
		0 & \frac{\dot{{f_x}}}{{f_x}} & 0 & 0 \\
		0 & 0 & \frac{\dot{{f_y}}}{{f_y}} & 0 \\
		0 & 0 & 0 & \frac{\dot{{f_z}}}{{f_z}} \\
		\end{array}
		\right)\cdot \left(
		\begin{array}{c}
		1 \\
		{f_x}\cdot x \\
		{f_y}\cdot y \\
		{f_z}\cdot z \\
		\end{array}
		\right)} \\
	& = \dot{\mathcal{E}}_{t}{\mathcal{E}_{t}}^{-1}\rho_t \equiv \mathcal{L}_t\rho_t.
	\end{align*}
	Here $\mathcal{L}_t$ would be indeterminate if $1/f_i$ were singular. In such a case, we seek well-defined higher-order derivatives to obtain a valid description for the evolution of states. Assuming that any function $f_i$ in the map has a zero at $t_{c}$ and supposing that $\dot{f}_i(t_c)$ is nonzero, then $\dot{f}_i/f_i$ does not exist at $t_{c}$. Here we can Taylor expand both $f_{i}(t)$ and $\dot{f}_{i}(t)$ around the critical time $t_{c}$ with $f_{i}(t_c) = 0$. If $\ddot{f}_{i}(t_c)$ is also zero and $\dot{f}_{i}(t_c)$ is nonzero, then 
	\[ \dfrac{\ddot{f}_{i}}{f_{i}} = \dfrac{\dddot{f_i}(t_c)}{\dot{f}_{i}(t_c)} \]
	is a well defined quantity at $t_c$ as well. Since ${\mathcal E}$ and ${\mathcal L}$ are both diagonal, it is straightforward to see that 
	\[ \frac{d^2 \rho}{dt^2} = {\mathcal L}_t^{(2)} \rho_t \]
	is a differential equation for $\rho(t)$ devoid of the singularities that beset the first-order equation. Here we have defined  higher-order generators as 
	\begin{equation}
	\label{higher_order_gen}
	\mathcal{L}^{(n)}_t \equiv \mathcal{E}^{(n)}_t\mathcal{E}^{-1}_t = \frac{d^n\mathcal{E}_t}{dt^n}\mathcal{E}^{-1}_t, 
	\end{equation}
	with ${\mathcal L}_t^{(1)} \equiv {\mathcal L}_t$. If the order of derivatives considered above does not lead to a nonsingular equation, we extend the same method to higher derivatives until we obtain a nonzero finite value for the ratio. 
	
	Note that this method may still not yield a finite value for some cases even if we consider all orders of derivatives. In such cases we find that a combination of different order generators with suitable weights of the form
	\begin{equation}
	\label{gen_eqn}
	\sum_n p_n \rho_t^{(n)} = 0
	\end{equation}  
	would yield a non-diverging time-local master equation that holds for all time. The coefficients $p_n$ can be obtained from the higher derivatives of the generator ${\mathcal L}_t^{(n)}$ as described in the next section. The exact dynamics can be found by solving these differential equations which require specifying more initial conditions than that for the traditional master equation. Our approach is valid for nondiagonal maps as well. For reasons of mathematical complexity and the lack of experimental literature requiring the usage of time-dependent Lindblad (or jump) operators, the singularities present in such dynamics are left unexplored in this paper.
	
	\section{\label{sec:Solution}Master equation for the spin model}
	The concept of higher-order equations can be nicely illustrated considering the example of the central spin model described earlier. It also offers a viable experimental setup to validate our findings. In general, characterizing the dynamics observed in an experiment requires an accurate description of the decay rates. Using the techniques of quantum process tomography, one may infer the relevant rates with sufficient accuracy as described below.
	
	In terms of traceless operators $F_{\alpha}$, Eq.~\eqref{GKSLeq} can be rewritten as
	\begin{align}
	\label{lindblad}
	\dot{\rho} =& -i [H(t),\, \rho_t] \nonumber \\  
	&+ \frac{1}{2} \sum_{\alpha,\, \beta=1}^{d^2-1}c_{\alpha\beta}(t)\left([F_{\alpha}\rho_t,\, F_{\beta}^{\dagger}] + [F_{\alpha},\, \rho_t F_{\beta}^\dagger]\right).
	\end{align}

	We choose $F_{\alpha}$ to be Pauli operators (upto a normalization constant) and $H = h_{\alpha}\sigma_{\alpha}$ is the Hamiltonian. Substituting this in Eq.~\eqref{lindblad} outputs a traceless matrix for $\dot{\rho}$. Since Pauli matrices form a basis for $2 \times 2$ matrices, we can express $\dot{\vec{\mathbf{r}}} \equiv (\dot{x},\,\dot{y},\,\dot{z})$ in terms of the nine Kossakowski coefficients $c_{\alpha\beta}$ and three parameters of the Hamiltonian.
	From the experimentally observed data, we can determine the values of $(\dot{x},\, \dot{y},\, \dot{z})$ at each time $t$ using
	$\dot{f} = \lim\limits_{h \rightarrow 0} [f(t+h)-f(t)]/h$ for $f \equiv (x,\, y,\, z)$.
	
	Corresponding to 12 unknowns (nine from $c_{\alpha\beta}$ and three more from $h_i$) and three known quantities ($\dot{x},\, \dot{y},\, \dot{z}$), we can setup 12 independent linear equations by choosing four linearly independent initial states. For example, the set of states $\rho_1 = \ket{0}\bra{0}, \, \rho_2 = \ket{1}\bra{1}, \, \rho_3 = \ket{+}\bra{+}$, and $\rho_4 = \ket{-}\bra{-}$ where $\ket{+} \equiv (\ket{0}+\ket{1})/\sqrt{2}$, and $\ket{-} \equiv (\ket{0}+i\ket{1})/\sqrt{2}$, furnishes one such choice.
	Determining all the unknowns involves solving the resulting linear equations. The first-order traditional master equation so obtained from the experimental data can now be used to locate the singular points.
		
	The quantum process tomography~\cite{Bellomo_2010_tomography,Bennink_2019_tomography, Boulant_2003_tomography,Howard_2006_tomography,Ben_2020_tomography} described above leads to the equation of motion given in Eq.~\eqref{spin_mod} and the corresponding dynamical map given in Eq.~\eqref{spin_map}. The generator of the dynamics is singular when one or more of the elements of the diagonal dynamical map goes to zero. By inspection, we see that these points correspond to the zeros of $\cos^N(\omega t)$ where $\omega \equiv 2A/\sqrt{N}$. As mentioned earlier, despite the singularities in ${\mathcal L}_t^{\rm spin} = \dot{\mathcal E}_t^{\rm spin} ({\mathcal E}_t^{\rm spin})^{-1}$, the dynamical map in Eq.~\eqref{spin_map} is analytic for all $t$. In order to construct a higher-order differential equation that avoids the singular behavior, we therefore start from the dynamical map $\rho_t = {\mathcal E}_t \rho_0$, where we have taken ${\mathcal E}_t^{\rm spin} \equiv {\mathcal E}_t$ for simplicity. We consider higher derivatives of the equation involving the dynamical map, 
	\begin{equation}
	\label{dyn_der}
	\rho_t^{(k)} = {\mathcal E}_t^{(k)} \rho_0,
	\end{equation}
	with the equation for $\rho_t^{(1)}$ being the same as Eq.~\eqref{spin_mod}. The strategy we adopt is as follows. The terms that appear in ${\mathcal E}_t^{(k)}$ are derivatives of $\cos^N(\omega t)$, which in turn are functions of $\sin(\omega t)$ and $\cos (\omega t)$. Computing a sufficient number of derivatives as in Eq.~\eqref{dyn_der} allows us to invert these functions and write them in terms of $\rho_t^{(k)}$ and the next suitable higher derivative of $\rho_t$ can be expressed fully in terms of its lower derivatives, leading to a higher-order dynamical equation of the form given in Eq.~\eqref{gen_eqn}. 
	
	The $x$ component for the Bloch vector representing $\rho_t$ is transformed by the dynamical map as $\rho_{t,x} = \cos^{N}(\omega t)\rho_{0,x}$. Since the $y$ component also follows the same pattern and since the map is diagonal, we focus on obtaining a higher-order differential equation for $\rho_{t,x}$ without loss of generality. The equation so obtained also applies to the full density matrix.  Exploiting the properties of derivatives of $\sin(\omega t)$ and $\cos(\omega t)$, we express any higher-order $\cos^{N}(\omega t)$ into a binomial sum of exponentials that upon simplification turns to a sum of cosines.
	\begin{equation}
	\cos^N(\omega t) = \frac{1}{2^N} \sum_{j=0}^{N}\binom{N}{j}e^{i(N-2j) \omega t}.
	\end{equation}
	For even $N$ we obtain a binomial sum of cosines as follows:
	\begin{equation}
	\cos^{2m}(\omega t) = \frac{1}{4^m} \binom{2m}{m} + \frac{1}{2^{2m-1}}\sum_{j=1}^m\binom{2m}{m+j}\cos(2j \omega t).
	\end{equation}
	Odd-order derivatives of $\rho_{t,x}$ contain $m$ terms each containing $\sin(2j\omega t)$ for $j=1, \ldots , m$. The first $m$ odd-order derivatives can be collected and rewritten as a system of linear equations of the form
	\[
	\begin{bmatrix}
	a_{11}   & \cdots & a_{1m} \\
	\cdots   & \cdots & \cdots \vphantom{\vdots} \\
	a_{m1}    & \cdots & a_{mm}
	\end{bmatrix}
	\begin{bmatrix}
	\sin(2 \omega t) \rho_{0,x}  \\ \vdots \\ \sin(2m \omega t)\rho_{0,x}
	\end{bmatrix}
	=
	\begin{bmatrix}
	\rho_{t,x}^{(1)}  \\ \vdots \\ \rho_{t,x}^{(2m-1)}  
	\end{bmatrix}
	\]
	where $a_{ij}$ denotes the coefficients gathered from odd differentiations,
	\begin{equation}
	a_{ij} = (-1)^{i} \frac{1}{2^{2m-1}} \binom{2m}{m+j} (2j \omega)^{2i-1}.
	\end{equation}

	The superscript on $\rho_{t,x}$ denotes the order of the time derivative. The binomial coefficient that appears in $a_{ij}$ is distinct and nonzero for each value of $j$, while the factor $(2j\omega)^{2i-1}$ is nonzero and different for each value of $i$ given a value of $j$. So we find that each $a_{ij}$ is non-zero and distinct which means that the determinant of the $m \times m$ matrix $A = [a_{ij}]$ is always nonzero. This system of linear equations can therefore be inverted so as to express $\sin(2j \omega t) \rho_{0,x}$ in terms of $d^{j}\rho_{t,x}/dt^{j}$ for $j=1,\ 3,\ \ldots,\ 2m-1$. The right hand side of the equation for the $(2m +1)$th derivative of $\rho_{t,x}$ is then completely determined by $\sin(2j \omega t) \rho_{0,x}$ for $j=1,\ldots m$, which in turn can be now written in terms of the odd derivatives of $\rho_{t,x}$. This leads to a differential equation of order $2m+1$ of the form $\sum_{j=0}^{m} p_{2j+1} \rho_t^{(2j+1)} = 0$. Here we have used the fact that both $\rho_{t,x}$ and $\rho_{t,y}$ have the same dynamics to write the differential equation for the full density matrix. 
	
	For odd $N$ we can do a similar analysis starting from 
	\begin{equation}
	\cos^{2m+1}(\omega t) = \frac{1}{2^{2m}}\sum_{j=0}^{m} \! \binom{2m+1}{j} \! \cos[(2m-2j+1)\omega t].
	\end{equation}
	The first $m+1$, odd-order derivatives $(\rho_{t,x}^{(1)}, \ldots, \rho_{t,x}^{(2m+1)})^T$ can be equated to
	\[
	\begin{bmatrix}
	a_{11}    & \cdots & a_{1,m+1} \\
	\cdots    & \cdots & \cdots \vphantom{\vdots} \\
	a_{m+1,1}   & \cdots & a_{m+1,m+1}
	\end{bmatrix}
	\begin{bmatrix}
	\sin( \omega t) \rho_{0,x} \\ \vdots \\ \sin[(2m+1) \omega t]\rho_{0,x}
	\end{bmatrix}
	\]
	with
	\begin{equation}
	a_{ij} = (-1)^{i} \frac{1}{2^{2m}}\binom{2m+1}{m+j} [(2j-1)\omega]^{2i-1}.
	\end{equation}
	This system of linear equations again yields $\sin[(2j+1) \omega t]$ for $j=0,\ 1,\	\ldots,\ 2m$ in terms of the odd order derivatives of $\rho_{t,x}$. Differentiating $\rho_{t,x}$ twice more leads to a master equation as desired. Note that when $N \rightarrow \infty$, $\cos^N(\omega t) \rightarrow e^{-2A^2t^2}$ with the singular behavior is pushed to $t \rightarrow \infty$. A Markovian, first-order, dephasing master equation is obtained in this limit with many states being mapped to the same state on the $z$ axis of the Bloch sphere asymptotically only.  
	
\begin{figure}[!ht]
		\resizebox{8.6cm}{5.9cm}{\includegraphics{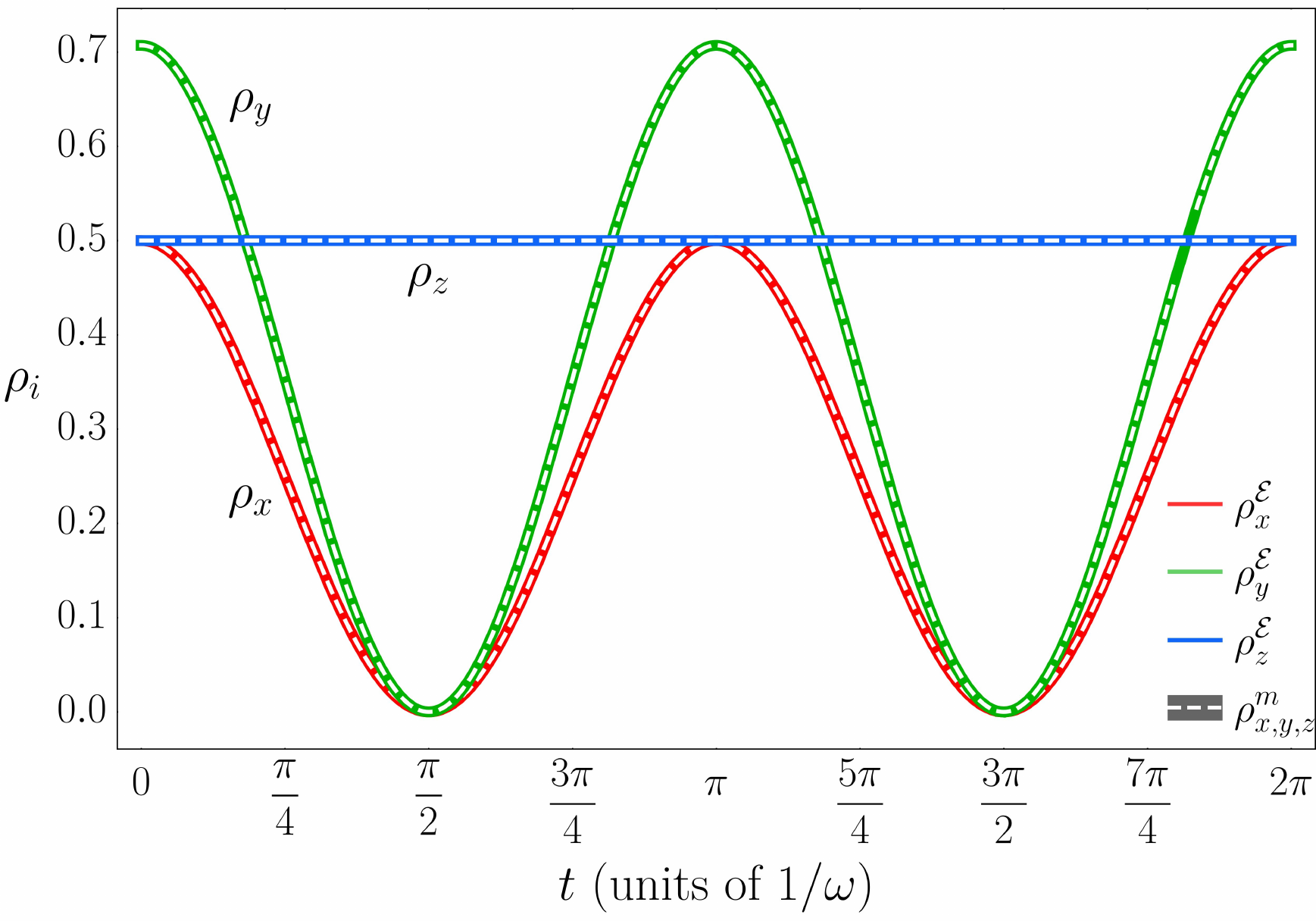}}
		\caption{\label{fig:num_soln} Numerical solutions of the higher-order master equation \eqref{spin_2} (dashed curves in white) are plotted component-wise along with the elements of dynamical map from Eq.~\eqref{spin_map} with $N=2$ (colored solid curves). This plot displays the evolution of each component of the Bloch vector for the initial pure state $\vec{r}_0 = \big( 1/2,\, 1/\sqrt{2},\, 1/2 \big)$. Solutions given by higher-order equations exactly agree with that of the dynamical map and the dashed white curves fall exactly on top of the solid colored ones. This is unlike the solution of the first-order equation which blows up at $\pi/2$ and cannot be propagated further. Time on the $x$-axis is shown in units of $1/\omega$.}
	\end{figure}

	For example, a third order master equation is obtained for $N=2$ and $\omega = 1$ in the central spin model. The corresponding dynamical map is $\mathcal{E}(t) = \diag\left(1,\,\cos^2(t),\, \cos^2(t),\, 1\right)$. Rewriting $\cos^2(t)$ as $[1+\cos(2t)]/2$ leads to $\dot{\rho}_t = -\sin(2t)\rho_0$, $\ddot{\rho}_t = -2\cos(2t)\rho_0$ and $\dddot{\rho_t} = +4\sin(2t)\rho_0$. Combining these derivatives we see that 
	\begin{equation}
	\label{spin_2}
	4\dot{\rho}_t + \dddot{\rho_t} = 0.
	\end{equation}
	This higher-order master equation for the central spin model with $N=2$ is numerically solved for a pure initial state $r_0 = \big( 1/2,\, 1/\sqrt{2},\, 1/2 \big)$ as shown in Fig.~\ref{fig:num_soln}. While the first-order equation~\eqref{spin_mod} is singular at $\pi/2$ and hence is unable to propagate the solution beyond that point, we see that the dynamics obtained from Eq.~\eqref{spin_2} is smooth at all times, just as desired.
	
	It is important to note that the higher-order equations can also be obtained by directly using the diverging generator and its derivatives leading to an equation that closely resembles Eq.~\eqref{gen_eqn}. We use $\rho_{t}^{(n)} = \mathcal{L}^{(n)}_t\rho_t$ in Eq.~\eqref{gen_eqn} so that $\sum_n p_n\mathcal{L}_t^{(n)}\rho_t = 0$ holds true for all $\rho_t$, which in turn yields,
	\begin{equation}
	\label{generator_equation}
	    \sum_{n} p_n \mathcal{L}^{(n)}_t = 0.
	\end{equation}
	For the central spin model it is possible to start from $\rho_t^{(1)} = {\mathcal L}_t \rho_t$ instead of Eq.~\eqref{dyn_der} and arrive at Eq.~\eqref{generator_equation} without considering the dynamical map. However the steps involved will be more complicated when using the generator rather than the map because of the $\rho_t$ appearing on the right hand side. While using Eq.~\eqref{dyn_der} makes it simpler to see how the higher-order equation is obtained, it also gives the impression that knowledge of the full dynamics in terms of the map at all times is necessary for obtaining the higher-order equation. We point out here that this is not the case and starting from the (singular) generator obtained using the process tomography steps outlined at the beginning of this section, one can directly obtain the higher-order master equation.
	
	We illustrate this approach for the central spin model with $N=2$ and $\omega=1$. As noted previously, the $x$ and $y$ components of Bloch vector undergo the same dynamics and so we consider only the $x$ component, $\rho_{t,x}$. Denoting the $x$ component of generator by $\mathcal{L}_{t,x}$, we have,
	\begin{align*}
	\mathcal{L}_{t,x} \rho_{t,x} &= -2 \tan(t) \rho_{t,x},\\
	\mathcal{L}^{(2)}_{t,x} \rho_{t,x} &\equiv \left(\dot{\mathcal{L}}_{t,x} + \mathcal{L}_{t,x}^2 \right) \rho_{t,x} = 2\left[\tan^2(t) - 1\right] \rho_{t,x}, \\
	\mathcal{L}^{(3)}_{t,x} \rho_{t,x} &\equiv \left(\ddot{\mathcal{L}}_{t,x} + 3\mathcal{L}_{t,x}\dot{\mathcal{L}}_{t,x} + \mathcal{L}_{t,x}^3 \right) \rho_{t,x} = 8\tan(t)\rho_{t,x}.
	\end{align*}
	From the equations above, as expected, we recover Eq.~\eqref{spin_2} in the form $4\mathcal{L}_t+\mathcal{L}^{(3)}_t = 0$.
	
	From either of the methods we described above to obtain the higher-order equations, it is clear that their order is $N+1$ for even $N$ and $N+2$ for odd $N$. Consequently, we would need as many specified initial conditions to overcome the issue of singularity. In other words, it is mandatory to know the history of the particle to determine the further evolution of a state. As this feature suggests the presence of memory effects to varying extents, it is then natural to speculate if a correspondence between the number of bath spins and the degree of non-Markovianity can be established. More on this is discussed in Sec.~\ref{sec:measures}.
	
	\section{\label{sec:more_ex} Higher order master equations for other types of singular open dynamics}
	
	The singular behavior for the first-order master equation of the central spin model is not unique to this model. We present several examples of CPTP maps with singularities, the first-order master equations, and their corresponding higher-order master equations whose solutions are free of singularities. As before, we phrase our discussion in terms of dynamical maps because of the simplicity and clarity afforded by this approach. Having the dynamical maps at hand also helps in verifying that the solutions of the higher-order master equations that are obtained indeed do reproduce the dynamics faithfully. We reiterate that as with the central spin model, the (singular) generator is sufficient to obtain the corresponding higher-order equations and knowledge of the full dynamics in terms of the dynamical map for all $t$ is not needed. We categorize the examples considered based on the dynamical map being unital or not.
	
	\subsection{Unital dynamical maps}
    We continue with the central spin model and consider a case where the locations of the singularities of the first-order master equation can be moved around by changing the model parameters. This means that the difficulties encountered in the numerical propagation of the first-order equation can be modulated and for certain choices of model parameters such solutions can become impracticable or even impossible to obtain. In this case, if one were to take the restricted point of view of an observer who has access only to the central spin and does process tomography to determine the form of the generator, the dynamical map for all times remains inaccessible to the observer since even numerical integration of the obtained first-order master equation may be precluded. Proceeding to construct the higher-order master equation then appears to be the only path forward for this restricted observer in order to gain predictive power over its evolution.
    
    We consider a central spin under the influence of two environment spins with unequal interaction strengths as given by the Hamiltonian,
	\begin{equation*}     
		H = \frac{\omega_{1}}{2} (\sigma_z \otimes \mathbb{I} \otimes \sigma_z) + \frac{\omega_{2}}{2}  (\sigma_z \otimes \sigma_z \otimes \mathbb{I}).
	\end{equation*} 
	and $\mathcal{E}_t = \diag(1,\, \cos( \omega_{1} t)\cos(\omega_{2} t),\,\cos( \omega_{1} t)\cos( \omega_{2} t),\, 1)$ is the map describing the reduced dynamics of the first qubit. The generator will include two tangent functions each with a different argument. It is possible to change the location of the singularity by altering the strength of interaction. In addition, if we increase the number of environment spins, the number of tangent functions in the generator will also increase. When singularities are aggregated, propagating the first-order differential equation beyond them, without accumulating significant errors becomes increasingly difficult. For the case of two environment spins, equations of motion for the $x$ and $y$ components of the Bloch vector of the state of the central spin are again the same (the dynamics of the $z$ component does not exhibit any singular behavior). The higher-order equation for $\rho_{t,x}$ is,
		\begin{equation}     
		\rho_{t,x}^{(4)} + 2(\omega_{1}^2+\omega_{2}^2) \rho_{t,x}^{(2)}+ (\omega_{1}^2-\omega_{2}^2)^2\rho_{t,x}=0.     
		\end{equation}
	Solving this fourth-order equation yields the correct dynamics.
	
	As a second example consider the dynamical map given below which is CPTP for all $\gamma, \omega \geq 0$ and has no inverse at $\omega t = (m+\frac{1}{2})\pi, m \in \mathbb{Z}$ due to the singular nature of the dynamics of the $x$ and $y$ components of the Bloch vector:
	\begin{equation}
	\label{ex1}
	\mathcal{E}_{t} = 
	\left(
	\begin{array}{cccc}
	1 & 0 & 0 & 0 \\
	0 & e^{-\gamma t} \cos (\omega t) & 0 & 0 \\
	0 & 0 & e^{-\gamma t} \cos (\omega t) & 0 \\
	0 & 0 & 0 & e^{-\gamma t} \\
	\end{array}
	\right).
	\end{equation}
	 The functions appearing in this dynamical map  are non-periodic and the singularities in the dynamics occur at periodic intervals of $\pi/\omega$.
	
	The traditional master equation for the above map is
	\begin{align}
	\dot{\rho}_t = \frac{1}{4}
	\bigg\{&\gamma (\sigma_x \rho_t \sigma_x - \rho_t) 
	+ \gamma (\sigma_y \rho_t \sigma_y - \rho_t) \nonumber \\
	+ &\left[\gamma + 2\omega \tan(\omega t)\right](\sigma_z \rho_t \sigma_z - \rho_t)\bigg\}.
	\end{align}
	For simplicity, assume that $\omega=\gamma=1$.
	The higher-order master equation for this example looks like 
	\begin{equation}
	\label{eq_ex2}
	\rho_t^{(4)}+M\rho_t = 0,
	\end{equation}
	where $M = \diag\left(0,\, 4,\, 4,\, -1\right).$
	Choosing different values of $\omega$ and $\gamma$ results in a master equation of different degree than the above.
	
	It would be misleading to dismiss the singularities in the first-order equations as manually avoidable by choosing to ``jump" over those discrete points while regularizing the traditional master equations, either by analytically integrating the rates or via forceful numerical techniques. Although one may try to ``escape" the singular points by carefully choosing the integration limits, it relies on having the exact knowledge of location of singularities. However, one can come up with examples where it is impossible to obtain all singular points analytically. The advantage of using higher-order equations is further emphasized by the fact that it is not necessary to know when singularities occur, as shown in the next example. Returning to the generic form in Eq.~\eqref{map_form} for the diagonal unital map, consider the following choice:
	\begin{subequations}
	\begin{align}
	\label{exp_decay}
	f_{x}(t) =  
	f_{y}(t) &= \frac{1}{6}(2+4e^{-\gamma t}-3\sin^2(\omega t)), \\  
	f_{z}(t) &= \frac{1}{3}(4e^{-\gamma t}-1).
	\end{align}
	\end{subequations}
	This dynamical map is constructed in such a way that it is not possible to obtain all the singular points analytically. In addition to $\gamma t = \log(4)$ and $\omega t = (m+1/2)\pi$ for $m = 0,\, 1,\, 2,\, \ldots$ for any $\gamma, \omega \geq 0$, the dynamics exhibits singular behavior whenever the following transcendental equation holds true:
	$\gamma t = \log 4 - \log(3 \sin^2(\omega t) - 2)$.
	The traditional master equation is
	\begin{align}
	\dot{\rho}_t = & \frac{\gamma}{4-e^{\gamma t}} 
	\bigg[\left(\sigma_x\rho_t\sigma_x-\rho_t\right) + \left(\sigma_y\rho_t\sigma_y-\rho_t\right)\bigg] \nonumber \\ 
	& + \left(\frac{\gamma}{e^{\gamma t}-4}+\frac{4\gamma + 3\omega e^{\gamma t}\sin(2\omega t)}{8+e^{\gamma t}[1+3\cos(2\omega t)]}\right) \nonumber \\ 
	& \times \left(\sigma_z\rho_t\sigma_z-\rho_t\right).
	\end{align}
	The higher-order equation provides a reliable description since it naturally gets rid of all the singularities regardless of our knowledge on their whereabouts. We obtain the following higher-order master equation when $\gamma = \omega = 1$ that holds for all times,
	\begin{align}
	\rho^{(5)}_t = 4\rho^{(1)}_t - 3\rho^{(3)}_t.
	\end{align}
	
	In this last example for unital maps, we demonstrate the presence of singularities due to the presence of zeros at discrete times, in all three diagonal elements of the dynamical map. For $1/n_1 + 1/n_2 + 1/n_3 \leq 1$ and $a_1,\, a_2,\, a_3 \geq 0$, the following choice of diagonal elements of the map from Eq.~\eqref{map_form} stays CPTP:
	\begin{subequations}
	\begin{align}
	f_{x}(t) &= 1-2\left(\frac{1-e^{-a_1t}}{n_1} + \frac{1-e^{-a_2t}}{n_2}\right),  \\ 
	f_{y}(t) &= 1-2\left(\frac{1-e^{-a_1t}}{n_2} + \frac{1-e^{-a_3t}}{n_3}\right),  \\ 
	f_{z}(t) &= 1-2\left(\frac{1-e^{-a_2t}}{n_2} + \frac{1-e^{-a_3t}}{n_3}\right).
	\end{align}
	\end{subequations}
	The constants $a_j$ and $n_j$ determine when the singularities occur and we can identify one set of singular points observed for each component of the Bloch vector of the state of the system qubit at times 
	\[ t_j=\frac{1}{a_j}\ln\left(\frac{1}{1-\frac{n_j}{4}}\right), \quad j=1,2,3.\]
	The rates appearing in the traditional master equation are given by \[ \gamma_x = \frac{1}{4}\left(\frac{\dot{f}_{x}}{f_{x}} - \frac{\dot{f}_{y}}{f_{y}} - \frac{\dot{f}_{z}}{{f_{z}}}\right)\] 
	and its cyclical permutations among $x, y,z$. For this map, singularities occur in all three Bloch vector components at distinct times determined by the constants $a_j$ and $n_j$. The higher-order equations without singularities that holds for all times is given by
	\begin{equation}
	M_{1} \rho^{(3)}_t + M_{2} \rho^{(2)}_t + M_{3} \rho^{(1)}_t = 0,
	\end{equation}
	where 
	\begin{eqnarray*}
		M_{1} & = &  \diag\left(0,\, 1,\, 1,\, 1\right), \\
		M_{2} & = & \diag\left(0,\, a_1+a_2,\, a_1+a_3,\, a_2+a_3\right), \\
		M_{3} & = &  \diag\left(0,\, a_1a_2,\, a_1a_3,\, a_2a_3\right).
	\end{eqnarray*}
	
	\subsection{\label{subsec:non_uni}Non-unital dynamical maps}
	The Jaynes Cummings Hamiltonian in the interaction picture for a two level atom coupled to a quantized electromagnetic field is given by
\begin{equation}
\hat{\mathcal{H}}_{\mathrm{JC}}= \omega\left(a \sigma_{+}+a^{\dagger} \sigma_{-}\right).
\end{equation}

	This model corresponds to a non-unital dynamical CPTP map~\cite{BreuerBook}:
    \begin{equation}
    \label{JC_map}
	\mathcal{E}_{\mathrm{JC}}(t) =
	\begin{bmatrix*}[c]
	1 & 0 & 0 & 0 \\
	0 & f(t) & 0 & 0 \\
	0 & 0 & f(t) & 0 \\
	f^2(t)-1 & 0 & 0 & f^2(t) \\
	\end{bmatrix*}.
	\end{equation}
	The corresponding generator is:
	\begin{equation}
	\label{JC_gen}
	\mathcal{L}_{\mathrm{JC}}(t) = 
	\begin{bmatrix*}[c]
	0 & 0 & 0 & 0 \\
	0 & \dot{f}/f & 0 & 0 \\
	0 & 0 & \dot{f}/f & 0 \\
	2\dot{f}/f & 0 & 0 & 2\dot{f}/f \\
	\end{bmatrix*}.
	\end{equation}
	This example is presented to show that our method can be applied to non-unital maps also. For a real function $f$, the time evolution corresponds to a time-local Lindblad-like master equation~\cite{Hall2007},
	\begin{equation}
	\label{JC_model}
	\dot{\rho}(t) = -\frac{\dot{f}(t)}{f(t)}\big[2 \sigma_- \rho_t \sigma_+ - \sigma_+\sigma_-\rho_t - \rho_t\sigma_+\sigma_-\big].
	\end{equation}
	where $\sigma_+ = \ket{e}\bra{g}$, and $\sigma_- = \ket{g}\bra{e}$. Choosing $f(t) = \cos(\omega t)$ corresponds to the Jaynes-Cummings model on resonance, describing the interaction of an atom with a cavity field. We see that Eq.~\eqref{JC_model} is singular just like Eq.~\eqref{spin_mod} because of the $\tan(\omega t)$ term. However, the regularized, higher-order master equation in this case will be different from the spin model ($N=1$) described earlier which has a second-order master equation. Noticing that
	\begin{equation}
	\label{JC_higher_gen}
	\mathcal{L}^{(4)} + 4\omega^2 \mathcal{L}^{(2)} = \diag\left(0,\, -3\omega^4,\, -3\omega^4,\, 0\right), 
	\end{equation}
	a straightforward calculation reveals that
	\begin{equation}
	\label{JC_corrected}
	M_1 \rho^{(4)}_t + M_2 \rho^{(2)}_t + M_3 \rho_t = 0,
	\end{equation}
	where
	\begin{align*}
	M_1 &= \diag\left(0,\, 1,\, 1,\, 1\right), \\
	M_2 &= \diag\left(0,\, 4\omega^2,\, 4\omega^2,\, 4\omega^2\right), \\
	M_3 &= \diag\left(0,\, 3\omega^4,\, 3\omega^4,\, 0\right).
	\end{align*}
    We can equivalently rewrite Eq.~\eqref{JC_higher_gen} as,	
	\begin{equation}
	\label{JC_higher_equiv}
	\rho^{(4)}_t+4\omega^2 \rho^{(2)}_t  = \frac{3}{2} \omega^4 (\sigma_z \rho_t \sigma_z - \rho_t). 
	\end{equation}
    The right hand side of Eq.~\eqref{JC_higher_equiv} has a different set of operators compared to Eq.~\eqref{JC_model} and it resembles a dephasing term with $\sigma_z$ operators rather than the $\sigma_\pm$ appearing in the first-order master equation. This highlights the fact that the higher-order master equations may have a substantially different form from the first-order ones in general. However, the presence of the higher derivatives means that these equations do not lend themselves to the usual interpretation of rates or Lindblad operators. For instance, in the present case, the operator $M_3$ acting on the state $\rho_t$ cannot be understood as a generator of time translations in the same manner as $\mathcal{L}_{\text{JC}}$. The meaning imparted by extra terms present in higher-order equations appear to be context-dependent and thus inferring their exact meaning is beyond the scope of this study. It may be noted that the dynamics described by both Eqs.~\eqref{JC_model} and \eqref{JC_higher_equiv} are the same except at the singular points.
	
	\section{\label{sec:measures}Comparing non-Markovian processes}
	
	The necessity to explore higher-order differential equations for a clear description of singular processes naturally begs the question of the relationship, if any, between the extent of non-Markovianity and the order of equations, or essentially, the nature of singularities. This prompts us to seek a means of comparing different singular non-Markovian processes using existing measures of non-Markovianity. Non-Markovianity manifests itself in various ways such that there is no single measure or a set of instructions by which comparison of its ``degree" can be conclusively done. Multiple measures have been developed as indicators of non-Markovian dynamics in the past, based on, for example, the nearest approximation to Markovian channels~\cite{Wolf2008}, entanglement between system and ancilla along with the deviations from the divisibility of dynamical maps~\cite{RHP2010}, non-monotonic behavior of fidelity~\cite{Vasile2011}, quantum Fischer information~\cite{Lu2010}, the volume of accessible states~\cite{Lorenzo2013}, non-zero quantum discord~\cite{Alipour2012}, and the behavior of trace distance~\cite{breuer_measure_2009, Laine2010, Liu2013}. There have been multiple studies to investigate the inflow of information and some of these studies have also considered those cases when the map is non-invertible~\cite{DivisibilityPRA2011,DivisibilityPRA2019,DivisibilityPRL2018}. Our interest, however, is on the relationship between nature of singularities and extent of non-Markovianity from the perspective of information inflow. We shall mainly focus on the trace distance measure defined in~\cite{breuer_measure_2009} owing to its quantitative nature and applicability to experimental realizations~\cite{Liu2011}.
	
	A quantum process is non-Markovian if there is an initial pair of states $\rho_1(0)$ and $\rho_2(0)$ such that the trace distance $D(\rho_1(t),\, \rho_2(t))$ starts to increase for some time $t > 0$. A measure of non-Markovianity introduced by Breuer, Laine and Piilo~\cite{breuer_measure_2009} defined in terms of this property is
	\begin{equation}
	\label{BLP}
	\mathcal{N}(\mathcal{E}_t) = \max\limits_{\rho_{1,2}(0)} \int_{t,\sigma>0}dt \ \sigma(\rho_1(0),\, \rho_2(0),\, t),
	\end{equation}
	where 
	\begin{equation} 	
	\label{rate_trace_dist}
	\sigma(\rho_1(0),\, \rho_2(0),\, t) = \frac{dD(\mathcal{E}_t\rho_1(0),\, \mathcal{E}_t\rho_2(0))}{dt},  
	\end{equation} 
	denotes the time derivative of the trace distance of the evolved pair of states. The trace distance for states $\rho_1$ and $\rho_2$, in turn is given by 
	\begin{equation}
	D(\rho_1,\, \rho_2) = \frac{1}{2}\text{Tr}\|{\rho_1-\rho_2}\|,
	\end{equation}
	where the modulus of an operator $A$ is $\|A\| = \sqrt{A^{\dagger}A}$.
	The integral over time in Eq.~\eqref{BLP} extends over all intervals in which $\sigma(t)>0$. The maximum is taken over all pairs of initial states $\rho_{1,2}(0)$. Note that the Breuer-Laine-Piilo (BLP) measure, $\mathcal{N}(\mathcal{E}_t)$ is a positive functional of the dynamical map $\mathcal{E}_t$ and that it acts as a measure for the maximal total inflow of information from the environment back to the open system. By construction, all Markovian processes have $\mathcal{N}(\mathcal{E}_t) = 0$. 
	
	For the spin model, $\sigma(t)$ is positive at periodic intervals and $\mathcal{N}(\mathcal{E}_t^{\text{spin}})$ adds up to infinity for any $N$ when the contributions from all the periods are added up. Therefore this measure cannot be used to compare the degree of non-Markovian behavior corresponding to different values of $N$. Analysis of other measures of non-Markovianity like the one quantified based on the change in Bloch sphere volume $V(t)$ of the set of accessible states of the evolved system~\cite{Lorenzo2013} also reveals a similar behavior independent of $N$ precluding the comparison that we seek. The divergent behavior of the BLP and related measures is not unique to the central spin model we consider. 

\begin{figure}[!ht]
		\resizebox{8.6cm}{5.9cm}{\includegraphics{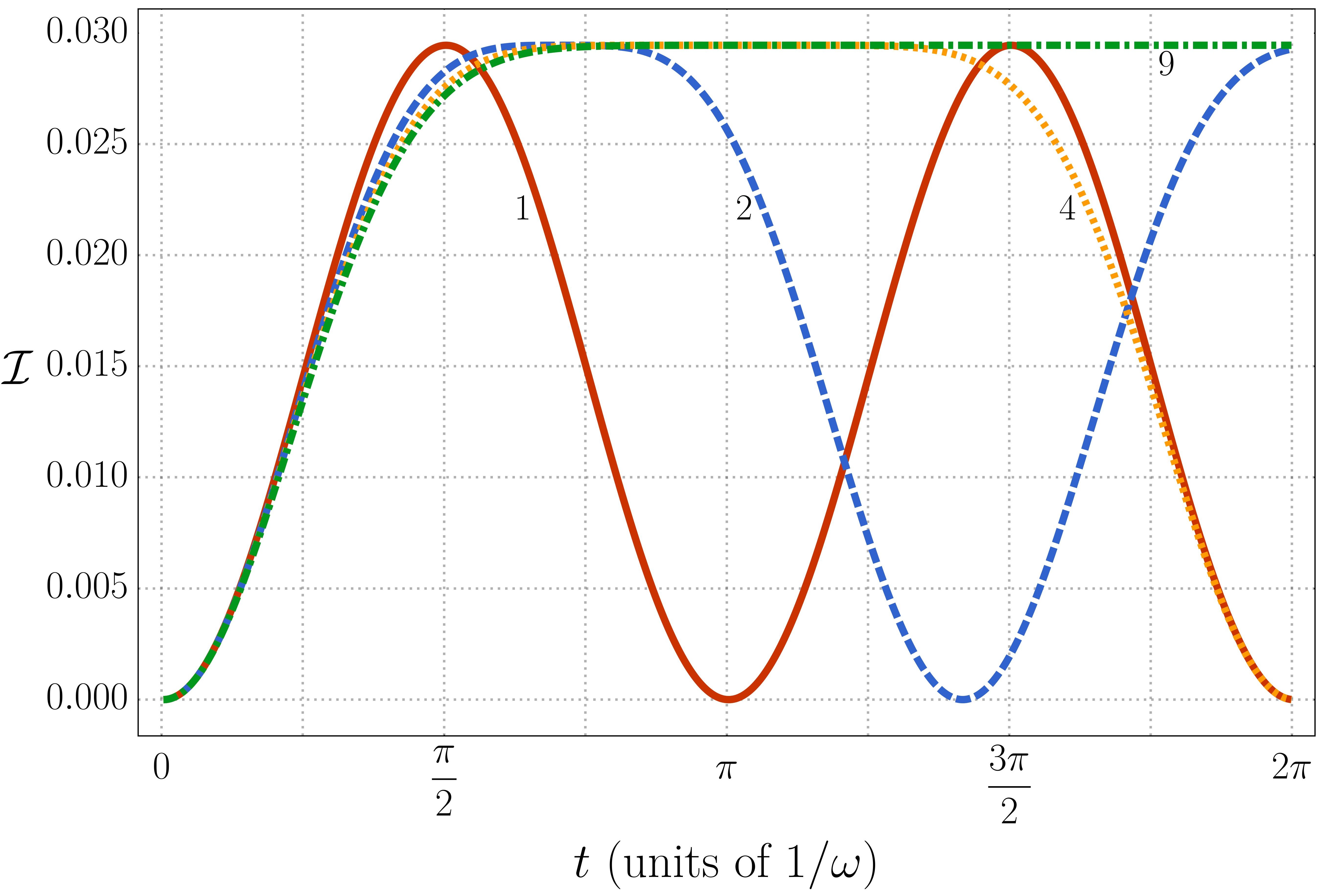}}
		\caption{\label{fig:mutinf}Mutual information between the central spin and environment is plotted for varying number of spins in the bath. The number of spins in the bath are placed as labels next to each curve in the plot. The higher the number of interacting spins, the longer the interval between vanishing of mutual information, eventually reaching infinity for large $N$. Here we have chosen $A = 0.5$ with $A_k = 0.5/\sqrt{N}$. Time on the $x$-axis is shown in units of $1/\omega$.}
	\end{figure}

	Information inflow from the environment to the system is an unmistakable signature of non-Markovian evolution. In order to explore the exchange of information of between the two in the central spin model, we look at the mutual information between the central spin and its environment of spins. Using the von Neumann entropy $S$ for a system $\rho$ calculated as $S(\rho) = -\text{Tr}(\rho \log \rho)$, the mutual information $\mathcal{I}$ is evaluated as
	\[
    \mathcal{I}(\rho_{\text{sys}}, \rho_{\text{env}}) = S(\rho_{\text{sys}}) + S(\rho_{\text{env}}) - S(\rho_{\text{joint}}),
	\]
	where $\rho_{\text{sys}}$ is the state of the system as in Eq.~\eqref{sys_state}, $\rho_{\text{env}} = \mathbb{I}/2^N$ is the bath state, and $\rho_{\text{joint}} = U\eta_0U^\dagger$ for $U = e^{-iH_{\text{spin}}t}$, all evaluated at time $t$.
	This mutual information is plotted for different values of $N$ in Fig.~\ref{fig:mutinf}.  We see from the oscillatory behavior of the mutual information that information is delocalized between the system and the environment and then localized back in the respective components in an alternating manner. The rate at which this exchange occurs depends on the number of environment spins, $N$. The time taken by the information, once delocalized, to again return to the central spin scales as $\sqrt{N}$. Note that this scaling is connected to the choice we made in Eq.~\eqref{spin_ham} for the Hamiltonian where the coupling between the central spin and the environment spins scaled as $1/\sqrt{N}$. We emphasize that this is  different from the example considered in Ref.~\cite{breuer_measure_2009} wherein the interaction Hamiltonian was not scaled with the number of spins in the environment. This choice resulted in a process that had no Markovian limit as a function of $N$. However, in our case, we recover the expected case of Markovian evolution as $N \to \infty$ with the delocalized information never returning to the central qubit.
	
	The dynamics of the mutual information highlights an aspect of non-Markovian evolution that is not typically addressed by the various known measures of non-Markovianity. While the amount of inflow of information from the environment is captured by a measure of non-Markovianity like the BLP measure, we see that central spin models with different $N$ are also characterized by the time scales at which the inflow happens. Non-Markovianity is indeed recognized as a feature that makes mathematical descriptions of physical phenomena rather difficult. In the absence of a comprehensive, all encompassing, understanding of non-Markovian quantum evolution, we are led to consider the possibility that more than one measure may be necessary for capturing different aspects of non-Markovianity. We consider whether persistence of information exchange is an aspect of non-Markovianity that can be quantified in a manner that it complements the existing measures. In addition to the central spin model, several processes allow the identification of `cycles' in their evolution such that the contribution of further dynamics to BLP measure after the first cycle is redundant. Taking a cue from this we propose supplementing the BLP measure with another quantity that determines a characteristic time $\tau$ over which the integral defining the BLP measure in Eq.~\eqref{BLP} can be limited to. The average rate of inflow of information over one such cycle can then be used as an effective quantifier that allows us to compare the degree on non-Markovianity of different processes belonging to the same family. In other words the ratio $\mathcal{N}(\mathcal{E}_t)/\tau$ with ${\mathcal N}({\mathcal E}_t)$ redefined as 
	\begin{equation}
	\label{BLP2}
	\mathcal{N}(\mathcal{E}_t) = \max\limits_{\rho_{1,2}(0)} \int_{t,\sigma>0}^\tau dt \ \sigma(\rho_1(0),\, \rho_2(0),\, t),
	\end{equation}
	becomes the figure-of-merit we explore in the subsequent discussion. 
	
	Finding an optimal pair of states that maximize the integral under consideration in Eq.~\eqref{BLP2} is made easier with the help of theorems proved in Ref.~\cite{Jyrki2012}, which state that an optimal pair of states must be orthogonal to each other and are restricted to the boundary of the state space. For qubit systems, this choice reduces to finding the optimal pair of pure, mutually orthogonal states that lie on the surface of the Bloch sphere. For all the examples discussed below, we have found the optimal pair of states by discretizing the surface of the Bloch sphere and evolving the antipodal states by the chosen dynamical map. The maximum of the sum of trace distances between evolved states over all the time intervals in $[0,\, \tau]$ for which $\sigma > 0$ is then divided by $\tau$ for determining the quantity of interest,
	\begin{equation}
	\mathcal{M}_{\tau}(\mathcal{E}_t) := \frac{\mathcal{N}(\mathcal{E}_t)}{\tau}.
    \end{equation}
    This rate of information inflow can be applied to any generic non-Markovian process.
    
	The purpose of cutoff time $\tau$ is to identify the time limit by which a pre-determined amount of information flows into the system from its environment. The interval $\tau$ varies greatly depending on the required proximity to the initial state. This statement is equivalent to choosing an error tolerance $\epsilon > 0$ for comparing the similarity of the dynamical map at a later time $\mathcal{E}_{t}$ with the initial map $\mathcal{E}_{0} = \mathbb{I}$. It is well known that in finite dimensional state spaces, all norms are equivalent~\cite{Conway1985}. Without loss of generality, we employ the $\mathcal{L}^{1}$-norm for measuring the distance between the dynamical maps. In other words, we need the first occurrence of time $\tau_\epsilon$ for which $\|\mathcal{E}_{\tau_\epsilon} - \mathcal{E}_{0}\|_{1} = 
	\sum_{i,j}|(\mathcal{E}_{\tau_\epsilon})_{ij} - (\mathcal{E}_{0})_{ij}| \leq \epsilon$, where $i$ and $j$ denote row and column indices, respectively, and $(d\mathcal{E}_t/dt)|_{\tau_\epsilon} < 0$ so as to select only those times for which the map is returning. Choosing a sufficiently smaller tolerance typically leads to longer recurrence times. Although the for purpose of comparing different non-Markovian processes belonging to the same family, the first occurrence of information inflow up to the prescribed tolerance level is sufficient, one might as well choose any such occurrence as long as comparisons are done on an equal footing.
	
\begin{figure}[!ht]
		\resizebox{8.4cm}{5.2cm}{\includegraphics{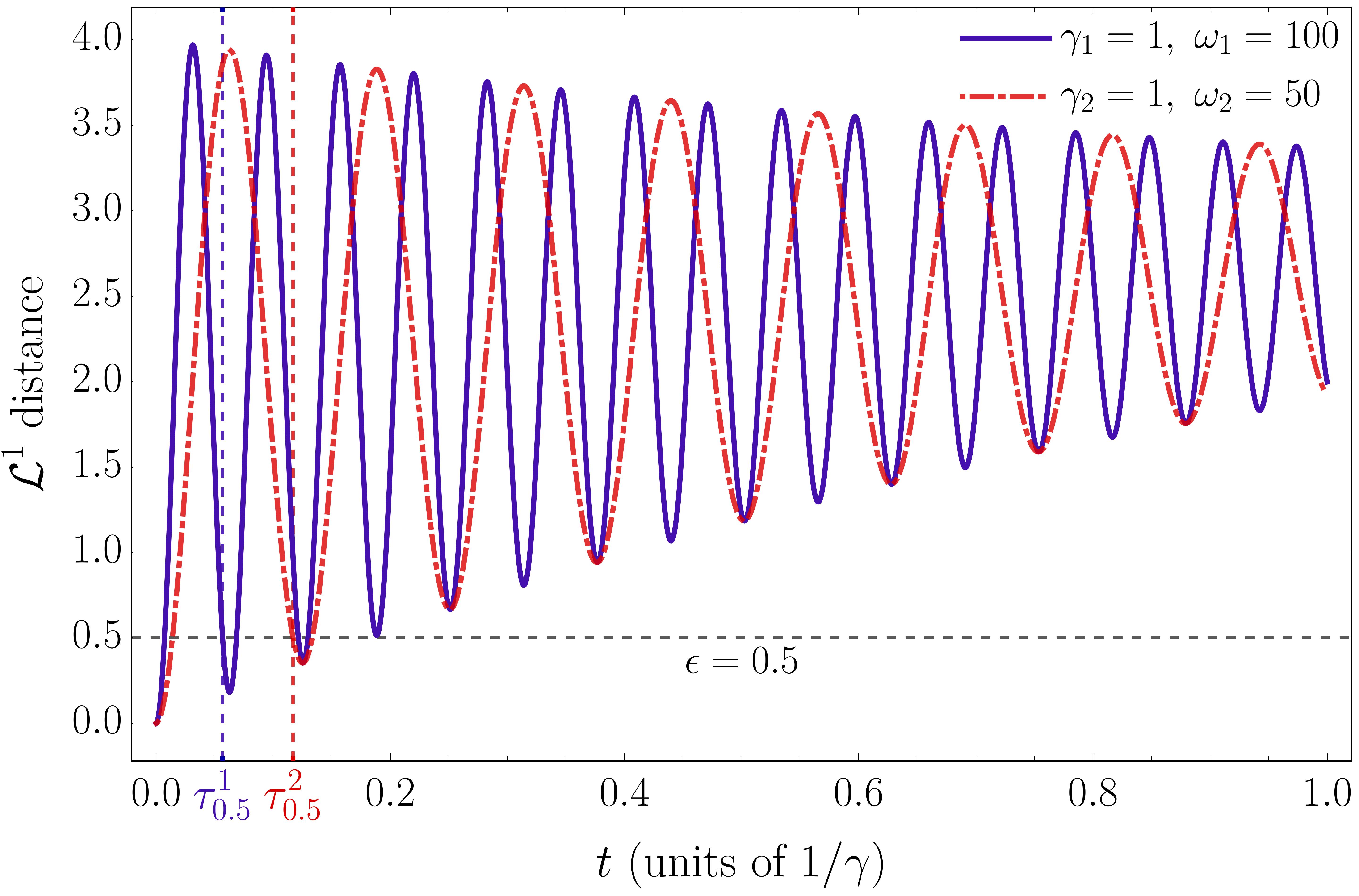}}
		\caption{\label{fig:non_periodic_case}$\mathcal{L}^1$ norm for the dynamical map from Eq.~\eqref{ex1} for the initial time and intermediate time is plotted as a function of time. The tolerance level is fixed at 0.5. The first arrivals of information inflow to the required tolerance are denoted by $\tau_{0.5}^{1}$ and $\tau_{0.5}^{2}$ for different oscillation frequencies, respectively. The lesser time for the recurrence of information inflow indicates a higher degree of non-Markovianity. Note that time has the units of $1/\gamma$ in this figure.}
	\end{figure}

	We will demonstrate the discussion above using the example described in Eq.~\eqref{ex1}, namely, $\mathcal{E}_t = \diag (1, \, e^{-\gamma t}\cos(\omega t),\, e^{-\gamma t}\cos(\omega t),\, e^{-\gamma t})$. 
	Consider two such processes with $\omega_1 = 100, \, \omega_2 = 50$ and $\gamma_1 = \gamma_2 = 1 \equiv \gamma$. Any general non-Markovian process, especially the ones with non-Markovian decay, need not bring the dynamical map as close to the identity matrix as desired and thus the tolerance level for comparison must be carefully chosen. We can mitigate this problem by choosing the first local minima for both the processes as the respective tolerance limits and then choose the maximum of the two to ensure both processes witness the norm reaching the assigned limits. For the processes at hand, we fix a tolerance level of $\epsilon = 0.5$. We desire to find the time $\tau$ for which $\|\mathcal{E}_\tau - \mathbb{I}\|_1 \leq 0.5$. We determine that $\tau_{0.5}$ is $0.0568$ and $0.1169$ for the first and second processes, respectively, as is evident from the Fig.~\ref{fig:non_periodic_case}. The quantity $\mathcal{M}_{\tau}^1$ for the first process turns out to be $30.1507$ and $\mathcal{M}_{\tau}^2$ is $14.3495$ for the second, which is consistent with the observation that the process having frequent oscillations turns out to be more non-Markovian than the one with slower oscillations.
	
	Defining a process-independent cutoff time $\tau$ for a non-periodic process is a challenging task. Hence, one may naively assign an infinite-time period for all such processes, allowing the BLP measure to also accumulate to infinity over an unbounded time interval. It is easy to see that $\mathcal{M}_{\tau}(\mathcal{E}_t)$ for any process is a bounded quantity. The key point is that the BLP measure is limited by the maximum difference in the trace distance for a pair of states and thus is always bounded by 1 for qubits. Since this increase in trace distance happens over a finite time, the proposed measure will always have a finite limiting value. However, it may not be straightforward to obtain the measure value in such cases.
	
    For the spin model, choosing tolerance limits as $\epsilon = 10^{-2}$ and $10^{-3}$ leads to cutoff times $\tau = 6.184$ and $6.252$, respectively, both of which are close to $2\pi$. The corresponding measure values turn out to be $0.3226$ and $0.3198$. One may as well choose $\epsilon = 0$ indicating complete inflow of information resulting in a cutoff time same as the period of the process which is $2\pi$. In addition to the generic procedure to find the cutoff time the $\tau$ for any process, the periodic and quasi-periodic processes offer simpler ways of fixing it.
    
	\textbf{Periodic Cases:}
	All periodic processes repeat their dynamics after their respective time periods $T$ and thus naturally furnish a time $\tau$ until which the BLP measure must be calculated. The complete dynamics of the system is captured by the dynamical map $\mathcal{E}_t$ whose period shall then ensure that all the states on the Bloch sphere revisit their initial configuration corresponding to $t = 0$ exactly and any dynamics beyond this period is redundant for eliciting the degree of non-Markovian behavior. Note that multiple pairs of states might revisit their initial configurations even before one cycle of the dynamical map is complete. By choosing the period of the map we are insisting that all states return to their positions in state space.  The initial configurations are typically ones in which there are no system-environment correlations, particularly if one considers only completely positive dynamical maps. Since all system states have reset their correlations, if any, with the environment at intervals defined by the period of the map, we can use $T$ as the upper limit of the integral in~Eq.~\eqref{BLP2}. The integral itself will have the same value if integrated over any interval of length $T$. The average rate of information inflow is then defined as, 
	\begin{equation}
	\label{measure}
	\mathcal{M}_\tau(\mathcal{E}_t) = \frac{1}{T} \max_{\rho_{1,2}(0)} \int\limits_{\substack{0 \\ \sigma > 0}}^{T} dt \, \sigma(\rho_1(0),\, \rho_2(0),\, t).
	\end{equation}
	
\begin{figure}[t]
		\resizebox{8.6cm}{5.5cm}{\includegraphics{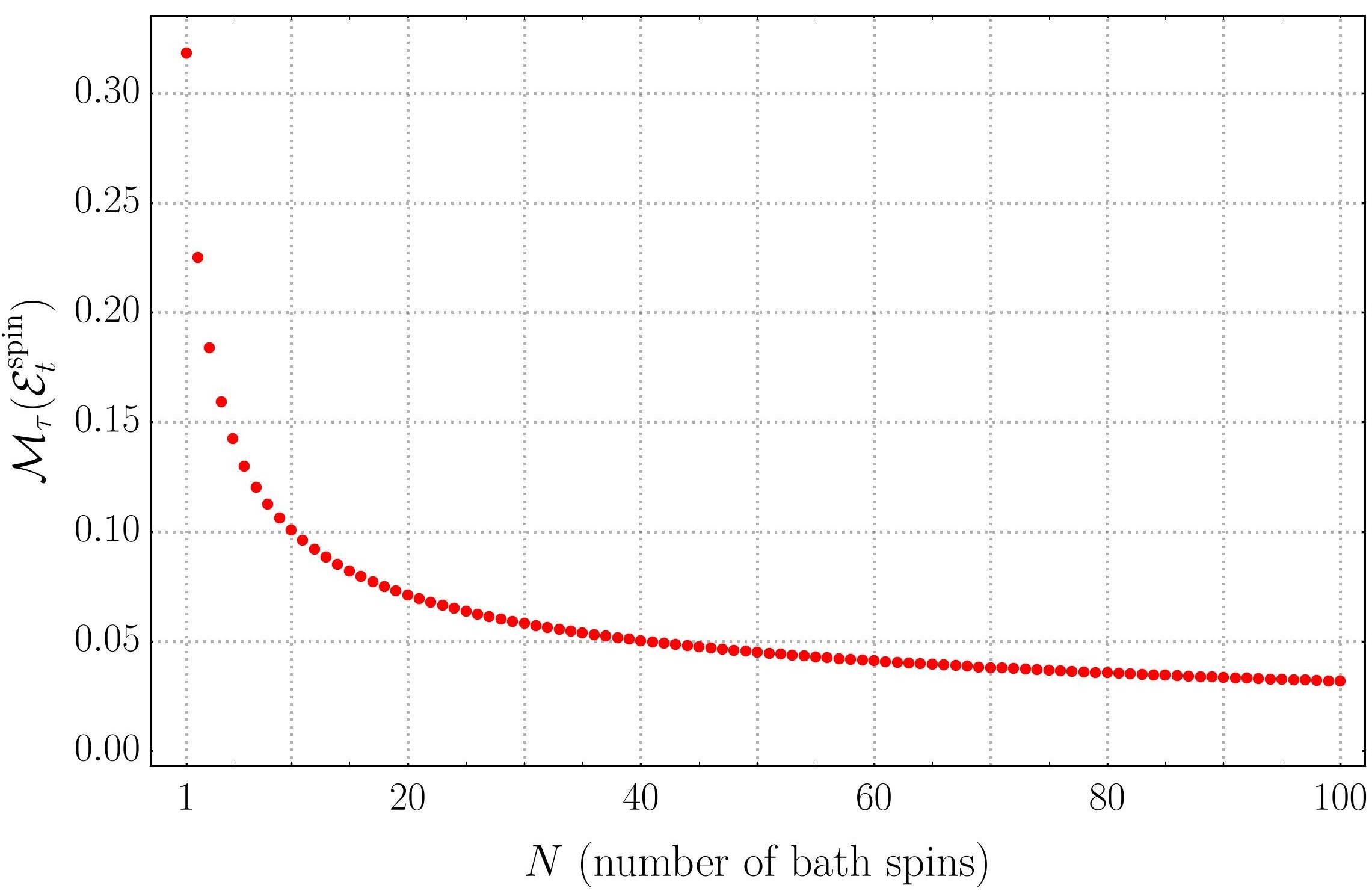}}
		\caption{\label{fig:blp_spin} Average rate of information inflow, $\mathcal{M}_\tau(\mathcal{E}^{\text{spin}}_t)$, for the central spin model plotted against the number of bath spins. Scaling the coupling constant in Eq.~\eqref{spin_ham} as $1/\sqrt{N}$ plays an important role in keeping the total interaction energy between the central spin and the environment constant, independent of $N$. In this case we see that the evolution becomes Markovian when $N \to \infty$ as expected with $\mathcal{M}_\tau(\mathcal{E}^{\text{spin}}_t)$ approaching zero in this limit.}
	\end{figure}

	We demonstrate the utility of Eq.~\eqref{measure} by applying it to the spin model described in Sec.~\ref{sec:spin_model}. Extension to other periodic cases is straightforward. For the spin model, the time period of the map depends on $N$. We find that $T = 2 \pi \sqrt{N}$ for odd $N$ and $\pi \sqrt{N}$ for even $N$ and $\sigma > 0$ in the interval $[\pi\sqrt{N}/2, \, \pi\sqrt{N}]$ for all $N$ and additionally in the interval $[3\pi\sqrt{N}/2, \, 2\pi\sqrt{N}]$ for odd $N$. The average rate of information inflow for this example is $\mathcal{M}_\tau(\mathcal{E}^{\text{spin}}_t) = 1/(\pi \sqrt{N})$. We see that $\mathcal{M}_\tau(\mathcal{E}^{\text{spin}}_t)$ is able to distinguish between central spin models with different number of bath spins and allow comparisons among them in terms of their degree of non-Markovianity. This is unlike the previously proposed measure of singular behavior from Ref.~\cite{Hou2012} where the value of the measure is $1/2$ irrespective of $N$. In our discussion of the dynamics of mutual information earlier, we noted that Markovian evolution is expected as $N \rightarrow \infty$. We see that as expected, $\mathcal{M}_\tau(\mathcal{E}^{\text{spin}}_t)$ converges to zero as $N$ becomes large as shown in Fig.~\ref{fig:blp_spin}, indicating Markovian limiting behavior. 
	
	We would like to highlight that the scaling constant directly affects the decay rate. Suppose the interaction strength in the Hamiltonian of the central spin model in Eq.~\eqref{spin_ham}is $B$. The corresponding average inflow rate from Eq.~\eqref{measure} is then proportional to $B$. In the discussion above, we have considered the interaction strength as $A/\sqrt{N}$ with $A=1/2$.
	
	\textbf{Quasi-periodic cases:}
	In what follows, we supplement our proposal for modification of BLP measure with an example where we find $\tau$ although the map has only an approximate periodicity. Consider the following 3-spin model with the Hamiltonian,
	\begin{equation*}
	 H = \frac{1+\pi}{4} (\sigma_z \otimes \mathbb{I} \otimes \sigma_z) +\frac{1-\pi}{4}  (\sigma_z \otimes \sigma_z \otimes \mathbb{I})
	\end{equation*}
	and the corresponding map describing the reduced dynamics of the first qubit, $\mathcal{E}_t = \diag(1,\,  [\cos(t) + \cos(\pi t)]/2,\, [\cos(t) + \cos(\pi t)]/2,\, 1)$. Clearly, there does not exist a period for this map since the two frequencies appearing (1 and $\pi$ in this case) are incommensurate. We propose two different approaches for finding the suitable time $\tau$.
	
	For all quasiperiodic processes, the general method can be understood as a corollary of the Poincar\'e recurrence theorem. The theorem states that for all finite dimensional systems with a time-independent Hamiltonian, the state vector $\ket{\psi(T)}$ returns arbitrarily close to the initial state $\ket{\psi(0)}$~\cite{Bocchieri1957}. Proceeding with the  method of obtaining $\tau$ earlier, we fix the error limit $\epsilon$ to $0.1$ for the three-spin example considered above which results in $\tau_{0.1} = 5.92$. Integrating Eq.~\eqref{measure} over all the intervals wherein the trace distance between a pair of states in increasing until $\tau_{0.1}$, we find the modified measure $\mathcal{M}_{\tau_{0.1}}$ to be $0.5204$. Similarly, $\tau_{0.01} = 43.95$ and $\tau_{0.001} = 43.98$ yield the measure values as $0.5130$ and $0.5128$, respectively. Noticeably, these values are more or less similar for different tolerance levels. Since higher accuracy can only be achieved after longer times, BLP measure values will also accumulate proportionally for the optimal pair of states. Thus we conjecture that the measure values remain almost the same for lesser tolerance values as well.
	
	The presence of quasi-periodicity allows us to adopt an alternative procedure which is as simple as rationalizing the irrational frequencies that appear so that the resulting terms of the dynamical map have a well-defined period. Since the irrationals are dense in $\mathbb{R}$, we are always guaranteed to find the rational approximation of any irrational number to the needed accuracy. For the case at hand, choosing $\frac{22}{7}$ as the approximation of $\pi$ yields the period as $14\pi$. The proposed measure $\mathcal{N}(\mathcal{E}_t)/\tau$ then has the value $0.5129$, which is also closer to the values obtained by the other method.
	
	In the sense of information inflow, we may conclude that certain processes are more non-Markovian than the others, as evidenced by the measure we introduced. In essence, the proposed addition to the BLP measure captures the differences in the degree of non-Markovianity between any two processes as advertised.
	
	\section{\label{sec:disc}Discussion and Conclusion}
	State preparation or initialization of a quantum system is a ubiquitous and important step in pretty much all experiments exploring the quantum realm. Initialization is an important step in running any algorithm in a quantum information processor and it is called for in most other applicable quantum technologies as well. Whether it is in the context of initializing an ensemble of identical quantum systems that are in different states into a common initial state or in the context of driving a single quantum system in an arbitrary state deterministically into a specific initial state, the preparation device has to induce dynamics on the system such that it is a many-to-one map of the kind we have discussed at length. During initialization, the quantum system of interest undergoes open quantum dynamics in contact with a preparation device that serves as its immediate environment. Our analysis shows that the preparation step can very well correspond to a singular point in the dynamics. Unless the strong assumption is made that after initialization the system and the preparation device are in a completely uncorrelated product state, further evolution of the system state may depend on the state from which the initialization process started. Note that in fact, the preparation device must return to the same quantum state after initialization irrespective of the system state that was prepared for all preparations to yield identical subsequent dynamics.

	In this paper we have explored in detail how such singular behavior in open quantum dynamics can be described mathematically using master equations with higher-order time derivatives. We see that such singular behavior may be much more common than previously imagined in the context of state preparations, lending added significance to our results. Our construction not only provides a means of propagating system states across the singular points of the normal first-order master equations, it also highlights the role that the environment can play in endowing various trajectories in state space that meet at the singular point with independent and distinct subsequent evolution. It may even be possible to observe subtle variations in subsequent trajectories of the same initial state in quantum process tomography experiments arising from differences in the starting point of state initialization and residual correlations that may exist between the system and state preparation device.
	
	It is interesting to note that from the various examples we have considered wherein higher-order master equations turned out to be useful, there is no particular discernible pattern for the structure of such equations. While a detailed characterization of the families of higher-order equations that may appear is beyond the scope of the present work, one way of understanding the possible origin of this variety is the following. Since the trajectories of multiple states coincide at the singular points, it is safe to say that at these points, relevant information that determines the future of each trajectory no longer resides in the state of the system. Given that quantum information can lie delocalized across multiple subsystems, this information can either lie delocalized across the system and its environment, be contained entirely in the state of the environment or both. The trajectories separating again can then be attributed to this information flowing back. The information inflow need not always produce a change in the state of the system that is first-order in time. It may affect higher-order time derivatives due to the interplay between the system-environment dynamics and the flow of information from the environment and/or from the delocalized form back to the system state.
	
	In a different context, this idea was presented in~\cite{Erika2012} where, using the Jaynes-Cummings model, an example was constructed in which two different system-environment interactions can lead to identical master equations but different trajectories for a qubit. The difference between the two Hamiltonians involves an instantaneous switch in the parameters that happens at a singular point in the dynamics such that the first-order master equation remains the same. However, two different solutions of the same master equation starting from the same initial state are obtained with and without the switch. From our point of view, the switch corresponds to rapidly changing the system-environment parameters exactly when the information that determines and differentiate subsequent dynamics of the qubit is not available in its state. The effect of the switch is in modifying higher-order time derivatives of the system state and so it does not appear in the first-order master equation. The trajectories followed by the same initial state with and without the switch are both solutions of the first-order equation but looking at the overall evolution it is easy to distinguish the two as expected.

	The role of information inflow from the environment into the system that disambiguates trajectories after singular points in the context of the experimentally implementable central spin model led us to the question of non-Markovian behavior in such models. With the aim of comparing the degree of non-Markovianity across different instances of the central spin model we introduced the typical time-scale for information inflow as a quantity that captures a different aspect of non-Markovian behavior compared to the standard approaches to quantifying such behavior. The average rate of information inflow introduced by combining this quantity with a well-established non-Markovianity measure helped us compare central spin models with different numbers of environment spins with regard to the degree of non-Markovianity in the evolution of the central spin. We also explored the limiting case of Markovian behavior that emerges when the number of environment spins become very large.  We then showed that the notion of an average rate of information inflow can be extended to generic non-Markovian open evolution as well and its applicability need not be limited to examples with singular behavior. We discussed these extensions for various types of non-Markovian dynamics possible for a single qubit.
	
	\begin{acknowledgments}
		The authors are thankful to Erika Andersson for directing attention to an important reference. Anil Shaji acknowledges the support of the SERB, Govt.~of India through grant no.~EMR/2016/007221 and the QuEST program of the Department of Science and Technology through project No.~Q113 under Theme~4. Jyrki Piilo acknowledges the support from Magnus Ehrnrooth Foundation. Vijay Pathak acknowledges the support of CSIR through fellowship (09/997(0040)/2015 EMR-I). Abhaya S. Hegde acknowledges the support of DST through INSPIRE fellowship (INSPIRE No. DST/INSPIRE-SHE/IISER-T/2008).
	\end{acknowledgments}
	\bibliography{ref.bib}
\end{document}